\documentclass[preprint,amssymb,aps,prb,showpacs,showkeys]{revtex4-1}

\usepackage[utf8x]{inputenc}
\usepackage[T1]{fontenc}
\usepackage{amsmath}
\usepackage{color}
\usepackage{graphicx}

\begin{document}
\date{\today}
\title{Influence of disorder on the structural phase transition and magnetic interactions in \(\mathrm{Ba}_{3-x}\mathrm{Sr}_x\mathrm{Cr_2O_8}\)}
\author{Henrik Grundmann}
\email{grundmann@physik.uzh.ch}
\author{Andreas Schilling}
\affiliation{Physik-Institut, Universit\"at Z\"urich, Winterthurerstrasse 190, CH-8057 Z\"urich}
\author{Marisa Medarde}
\affiliation{Laboratory for Development and Methods, Paul Scherrer Institut, 5232 Villigen PSI, Switzerland}
\author{Denis Sheptyakov}
\affiliation{Laboratory for Neutron Scattering, Paul Scherrer Institut, 5232 Villigen PSI, Switzerland}

\begin{abstract}
\noindent 
The spin dimer system \(\mathrm{Ba}_{3-x}\mathrm{Sr}_x\mathrm{Cr_2O_8}\) is a solid solution of the triplon Bose-Einstein condensation candidates \(\mathrm{Ba_3Cr_2O_8}\) and \(\mathrm{Sr_3Cr_2O_8}\). The magnetic intradimer interaction constant \(J_0\) in this spin system can be tuned by varying the Sr content \(x\). Very interestingly, this variation of \(J_0\) with \(x\) is highly nonlinear. In the present study, we show that this peculiar behavior of \(J_0\) can be only partly explained by the changes in the average crystal structure alone. We report on neutron powder diffraction experiments to probe the corresponding structural details. Performing extended Hückel tight binding calculations based on those structural details obtained at liquid helium temperatures, we found that the change of the magnetic interaction constant can be well reproduced by taking into account the presence of a structural transition due to the Jahn-Teller active Cr\(^{5+}\)-ions. This transition, lifting the orbital degeneracy and thereby the magnetic frustration in the system, is heavily influenced by disorder in the system arising from partially exchanging Ba with Sr.
\end{abstract}
\maketitle
\section{Introduction}
Bose-Einstein-condensation (BEC) is one of the most fascinating aspects in modern physics. A decade ago, the field of spin dimer physics has been closely linked to this topic by explaining several properties of certain spin dimer systems in terms of a BEC of magnetic quasiparticles (triplons)\cite{nikuni_tlcucl3_magnetization}. The triplons, formed by dimers of magnetic ions, condense at a low enough temperature \(T\) and above a critical magnetic field \(H_c(T)\). Once the condensate is formed, the triplons should show macroscopic phase coherence, a defining property of every BEC. It has very recently been suggested that this phase coherence could be probed in a device of two coupled spin dimer materials with different critical fields\cite{schilling_josephson}, which would represent a direct analogue to a Josephson junction showing the a.c.-Josephson effect. As in superconductors and atomic BEC's\cite{dc_bec_josephson,acdc_bec_josephson}, this effect is based on the existence of a phase coherent macroscopic wave function. A successful detection of Josephson effects in spin dimer systems would therefore be a direct proof for macroscopic phase coherence and allow to classify the observed spin condensate as a true BEC.

As the above sketched  experiment requires two spin systems with a certain difference in the respective critical fields \(H_c\), the fine tuning of this parameter becomes necessary. The critical field largely depends on the intradimer magnetic interaction constant \(J_0\) between the ions that form the spin dimers. The \(J_0\) is essentially determined by the specific \mbox{(super-)exchange} path between those ions and therefore changes if the crystal structure is modified. The spins systems \(\mathrm{Ba_3Cr_2O_8}\) and \(\mathrm{Sr_3Cr_2O_8}\), two candidates for a triplon BEC\cite{aczel_ba3cr2o8,aczel_sr3cr2o8,zvyagin_ba3cr2o8}, represent such spin systems with a similar structure, but strongly differing magnetic interactions. For the corresponding mixed system \(\mathrm{Ba}_{3-x}\mathrm{Sr}_x\mathrm{Cr_2O_8}\), \(J_0\) has been reported to be tunable by changing the Sr content \(x\)\cite{mrbpaper}. It was found that \(J_0\) changes in a peculiar, non-monotonous way as a function of \(x\) which contrasts with the almost linear changes of the lattice parameters. Up to now, this discrepancy has not been resolved.

\begin{figure}
 \includegraphics{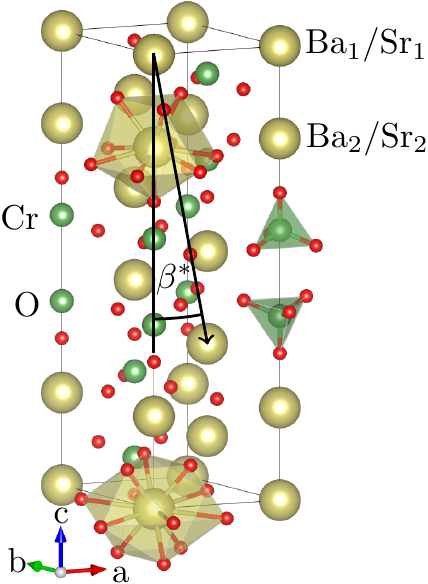}
 \caption{Crystal structure of \(\mathrm{Ba}_{3-x}\mathrm{Sr}_x\mathrm{Cr_2O_8}\) at room temperature. The Cr-tetrahedron is shown in dimer configuration on the right, the Ba\(_1\)/Sr\(_1\)-dodecatope is shown on the bottom and the Ba/Sr\(_2\)-decatope is shown on the top of the drawing. The angle \(\beta^*=\beta-90^\circ\) is marked to show how the monoclinic angle \(\beta\) was calculated for Fig. \ref{fig:shiftfig}.}
 \label{fig:strukturfig}
\end{figure}

The present work aims to give a satisfactory explanation based on a more detailed examination of changes in the crystal structure with \(x\). It has been shown for \(\mathrm{Sr_3Cr_2O_8}\) that due to a Jahn-Teller type, temperature induced structural phase transition involving certain oxygen positions (which is present for both \(\mathrm{Ba_3Cr_2O_8}\) and \(\mathrm{Sr_3Cr_2O_8}\)), \(J_0\) is strongly enhanced at at low temperatures\cite{Chapon}. Using neutron powder diffraction techniques which are more sensitive to the oxygen positions than corresponding X-ray diffraction data in the presence of heavy elements such as Sr and Ba, we have investigated the crystal structure of \(\mathrm{Ba}_{3-x}\mathrm{Sr}_x\mathrm{Cr_2O_8}\) for several values of \(x\) at room temperature and at \(T=2\,\mathrm{K}\). Based on the obtained lattice parameters and atomic positions, we performed extended Hückel tight binding (EHTB) calculations to estimate \(J_0\) as a function of \(x\).

\section{Experimental Details}
\subsection*{Synthesis}
 The polycrystalline samples were prepared by standard solid state reaction schemes. Powders of \(\mathrm{BaCO_3}\) (99,98\%, \emph{Sigma-Aldrich}), \(\mathrm{SrCO_3}\) (99,9\%, \emph{Sigma-Aldrich}) and \(\mathrm{Cr_2O_3}\) (99,9\%, \emph{Sigma-Aldrich}) were mixed according to \begin{equation*}(3-x)\mathrm{BaCO_3} + x\mathrm{SrCO_3 + Cr_2O_3\rightarrow Ba}_{3-x}\mathrm{Sr}_x\mathrm{Cr_2O_8+CO_n},\end{equation*} ground and heated in flowing Ar at \(1300\,^\circ\mathrm{C}\). The heating was started with a linear ramp up to \(1300\,^\circ\mathrm{C}\) in \(5\,\mathrm{h}\). The samples remained at this temperature for \(12\,\mathrm{h}\), followed by a linear cooling to room temperature in  \(8\,\mathrm{h}\). The grinding and heating  was repeated twice. We prepared samples with \(x\in\left\{0,\frac13,\frac23,1,\frac43,\frac53,2,\frac73,\frac83,3\right\}\).
\subsection{Neutron Diffraction}
The neutron powder diffraction experiments were carried out at the high resolution powder diffractometer HRPT\cite{Fischer_HRPT} at Paul-Scherrer-Institute (PSI, Villigen) at room temperature and at \(T=2\,\mathrm{K}\), using \(\lambda=1.494\,\textup{\AA}\) [Ge(533)]. Scans were performed with fixed primary soller collimation 
(40') and a secondary slit of (40'). The powder samples (\(\approx6\,\mathrm{g}\)) were loaded into cylindrical vanadium containers of \(8\,
\mathrm{mm}\) diameter (\(6\,\mathrm{mm}\) for \(\mathrm{Ba_\frac83Sr_\frac13Cr_2O_8}\)). The low temperature measurements were performed using a He 
cryostat, whose contribution to the total scattering was minimized using an oscillating radial collimator. The obtained diffractograms were analyzed 
using the Rietveld package \emph{FullProf} suite (available free of charge at http://www.ill.eu/sites/fullprof/.).
\subsection*{Tight binding calculations}
\begin{figure}
 \includegraphics{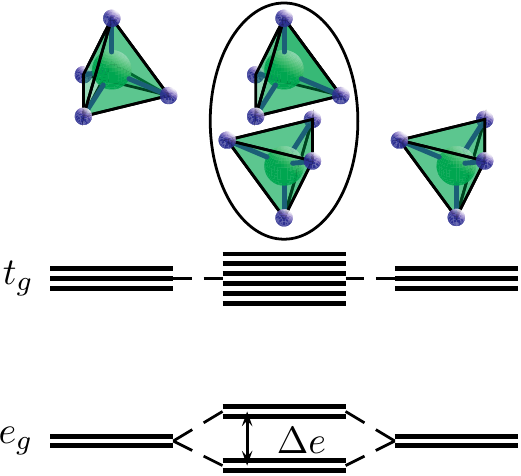} 
 \caption{Sketch for the orbital splitting \(\Delta e\) that occurs when two CrO\(_4^{3-}\)-tetrahedra form a dimer. The sketch refers to the undistorted room temperature structure.}
 \label{fig:splitfig}
\end{figure}

Based on these structure data, we calculated the antiferromagnetic interaction constant \(J_0\)  using the extended Hückel tight binding method which was successfully used in the past to estimate the interaction constants in Ba\(_3\)Cr\(_2\)O\(_8\) \cite{whangbo_ba3cr2o8} and Sr\(_3\)Cr\(_2\)O\(_8\) \cite{Chapon}. These calculations are based on determining the orbital splittings \(\Delta e_{\mu\mu}\) of the occupied chromium orbitals. This splitting \(\Delta e_{\mu\mu}\) of the \(\mu\)-th considered orbital occurs when two CrO\(_4^{6-}\)-tetrahedra are brought together to form a dimer (see  Fig. \ref{fig:splitfig}) and it can be calculated directly. Based on these splittings, \(J_0\) can be estimated as
\(J_0=\frac{\langle (\Delta e)^2\rangle}{U}\), where \(U\) is a repulsion potential and \(\langle (\Delta e)^2\rangle=\frac1n^2\sum\limits_\mu(\Delta e_{\mu\mu})^2\) is a sum over all \(n\) occupied orbitals. Due to the degeneracy of the \(3z^2-r^2\)- and the \(x^2-y^2\)-orbital at room temperature, both orbitals have to be taken into account  when calculating \(\langle (\Delta e)^2\rangle\). At temperatures below the Jahn-Teller transition, however, only the \(3z^2-r^2\) orbital is occupied (see below), so that \(\langle (\Delta e)^2\rangle=(\Delta e_{3z^2-r^2})^2\). A more detailed description of the general implementation of this procedure and its limitation to antiferromagnetical interactions can be found in \cite{whangbo_review}. The ETHB calculations have been performed using the \emph{SAMOA}-suite (available free of charge at http://www.primec.com/‎.).

The repulsion potential \(U\) depends on the chemical composition and should vary gradually with the Sr concentration \(x\)\cite{dimer_interaction}. Assuming a linear behavior as a function of \(x\), \(U(x)=A\cdot x+B\), we calculated \(J_0(x)\) from \(J_0=\frac{\langle \Delta e\rangle^2}{A\cdot x+B}\). Values for \(A\) and \(B\) were determined by comparing the published experimental values of \(J_0\) (from inelastic neutron scattering)\cite{kofu,quinteracastro_sr3cr2o8} with the calculated orbital splitting for the considered compounds in the low temperature phase.\\

\section{Results and Discussion}
\subsection*{Crystal structure at room temperature}
\begin{figure}
 \includegraphics{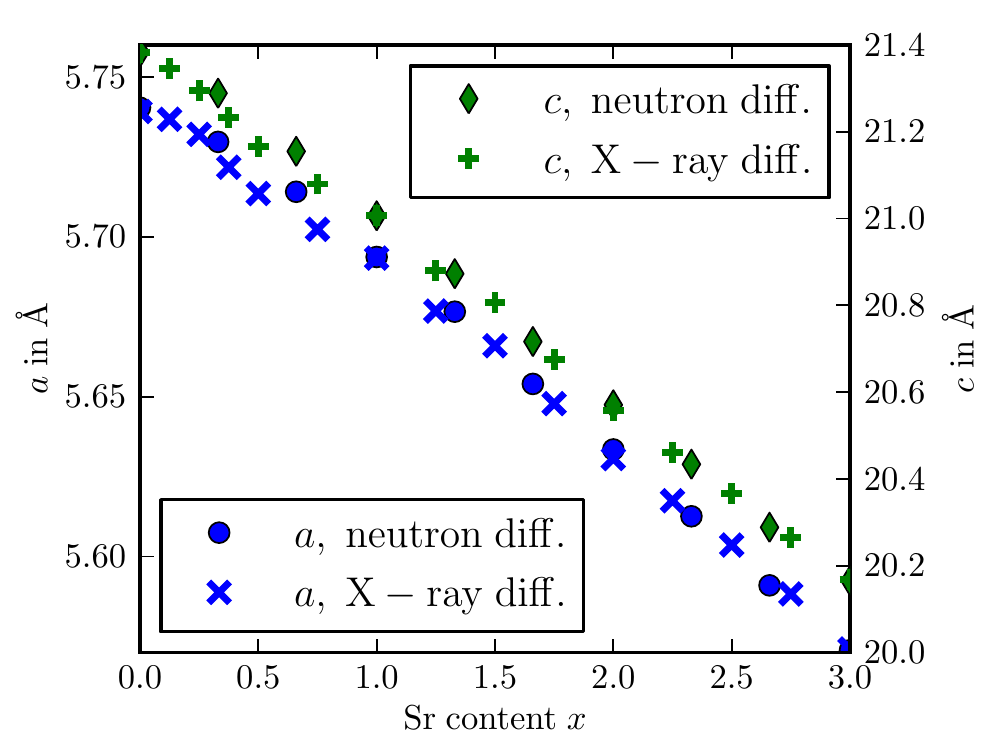}
 \caption{Lattice constants \(a\) and \(c\) of \(\mathrm{Ba}_{3-x}\mathrm{Sr}_x\mathrm{Cr_2O_8}\) at room temperature (space group R\(\overline{3}\)m). The lattice constants are shown as a function of the Sr content \(x\) as obtained from X-ray\cite{mrbpaper} and neutron powder diffraction experiments. The error bars for the values obtained from neutron diffraction are smaller than the size of the markers.}
 \label{fig:latcon}
\end{figure}

In the neutron powder diffraction patterns obtained at room temperature, the observed Bragg reflections could be indexed using the space group R\(\overline{3}\)m for all \(x\). From the Rietveld refinement of the diffraction data we obtained an almost linear behavior of the lattice constants \(a\) and \(c\) as a function of the Sr content, in accordance with recently reported X-ray diffraction experiments (see  Fig. \ref{fig:latcon}). 

\begin{figure}
 \includegraphics{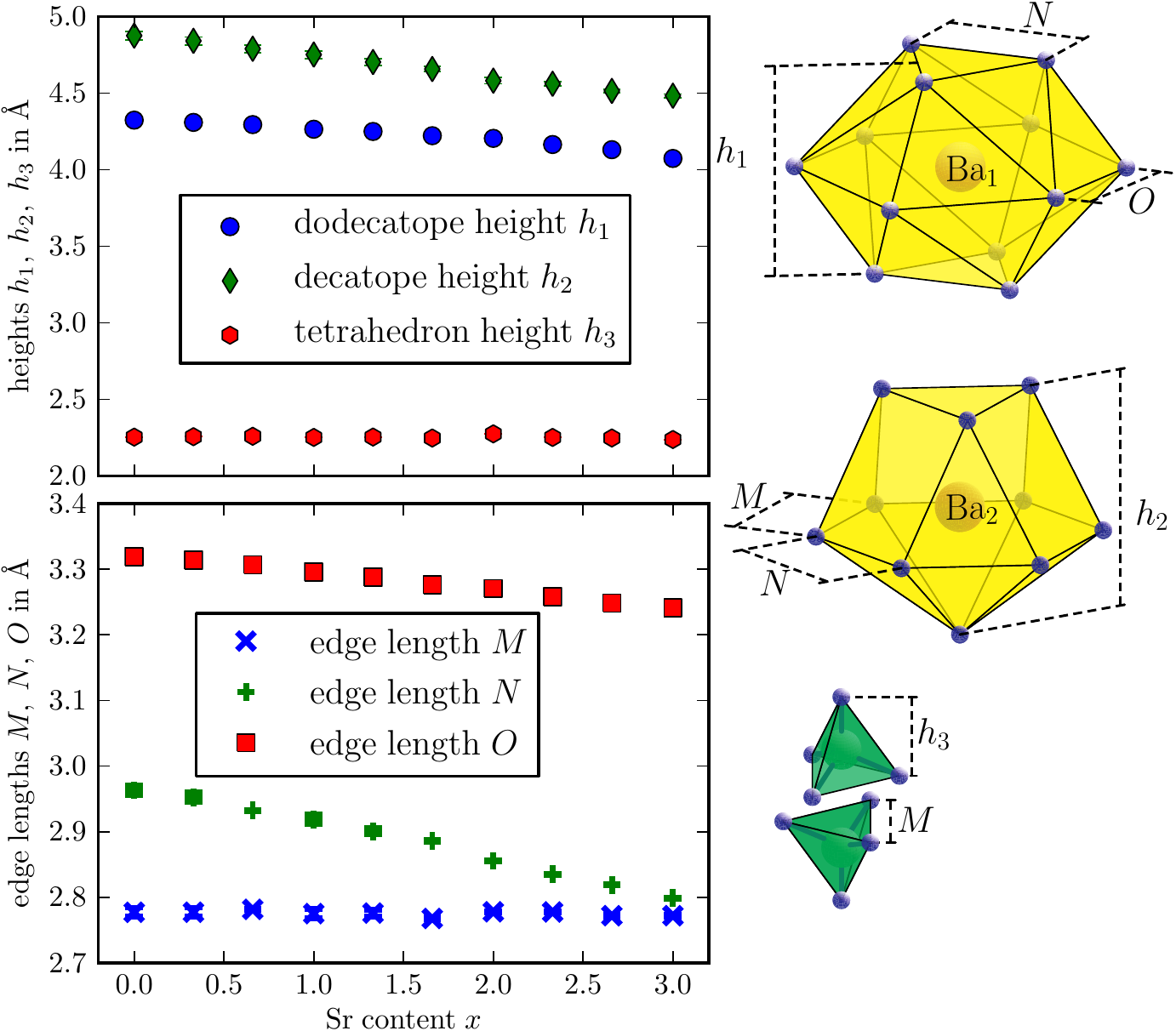}
 \caption{Height and edge length of the oxygen polyhedra in \(\mathrm{Ba}_{3-x}\mathrm{Sr}_x\mathrm{Cr_2O_8}\) as functions of the Sr content \(x\) at room temperature. The sketches on the right show the definitions of the plotted parameters.}
 \label{fig:dimfig}
\end{figure}

Along with this almost linear decrease of the lattice constants, the size of the oxygen polyhedra that surround the Ba/Sr-ions decreases linearly as well. The Ba/Sr atoms are located at two different positions (sites Ba\(_1\) and Ba\(_2\)). The respective Wyckoff positions are \(3a\) and \(6c\) for R\(\overline{3}\)m (\(4e\) and \(8f\) for \(C_{2/c}\)). The atoms located in the Ba\(_2\)-position with point site symmetry \(3m\) are 10-fold coordinated.  The height \(h_2\) of the coordinating decatope changes linearly as a function of \(x\) (see Fig. \ref{fig:dimfig}). The planar oxygen hexagon surrounding the Ba\(_2\) position is not evenly formed. Although the inner angle is always \(\alpha=120^\circ\), the edge \(M\) that is shared with the chromium tetrahedron is shorter than the edge \(N\) that is shared with the Ba\(_1\) dodecatope. The value of \(N\) decreases with increasing \(x\) until \(M\) and \(N\) are equal for \(x=3\) (see  Fig. \ref{fig:dimfig}). 

The position Ba\(_1\) is twelve-fold coordinated with site symmetry \(\overline{3}m\) (see  Fig. \ref{fig:dimfig} a). As stated above, the edge length \(N\) is shared with the Ba\(_2\)-decatope. Both edge lengths \(N\) and \(O\) as well as the dodecatope height \(h_1\) decrease linearly as a function of \(x\) (see Fig. \ref{fig:dimfig}).

The oxygen tetrahedron around the Cr-ion is only slightly affected by varying the stoichiometry. Neither the tetrahedron height \(h_3\) nor the edge length \(M\) of the base triangle change significantly as a function of \(x\) (see Fig. \ref{fig:dimfig}). Without any additional influences, this negligible change of the Cr-tetrahedron should be reflected in a rather small, monotonous change of the interaction constant. This change of \(J_0\) would be mostly given by the shrinking separation of the two tetrahedra forming a dimer as a function of \(x\). Thus, the structural data at room temperature does not give a satisfactory explanation for the observed changes of \(J_0\) with the stoichiometry.

\subsection*{Crystal structure at \(T=2\,K\)}
\begin{figure}
 \includegraphics{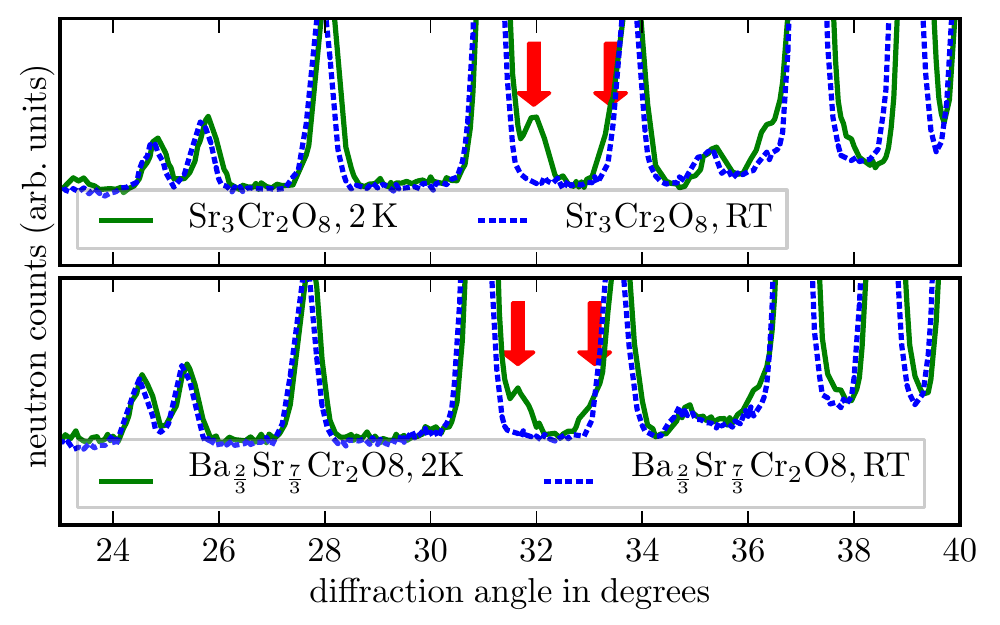} 
 \caption{Additional peaks appearing in the low temperature neutron diffractograms for \(\mathrm{Sr_3Cr_2O_8}\) and \(\mathrm{Ba_\frac23Sr_\frac73Cr_2O_8}\), indicating the occurence of the structural phase transition.}
 \label{fig:breakfig}
\end{figure}

Our analysis of the diffraction patterns obtained at \(T=2\,\mathrm{K}\) shows that the lattice symmetry is lowered for some, but not all values of \(x\) upon cooling. For chemical compositions close to the parent compounds, i.e. for \(x\in\{0,0.33,2.33,3\}\), the lattice is better described using the space group \(C_{2/c}\), that has been reported for Ba\(_3\)Cr\(_2\)O\(_8\) and Sr\(_3\)Cr\(_2\)O\(_8\), with new diffraction peaks appearing (see  Fig. \ref{fig:breakfig}). For the remaining samples, no superstructure peaks could be observed. The diffraction pattern at \(T=2\,\mathrm{K}\) could be well described by the room temperature space group R\(\overline{3}\)m, indicating a suppression of the structural phase transition for the samples with \(0.33<x<2.33\). 

\begin{figure}
 \includegraphics{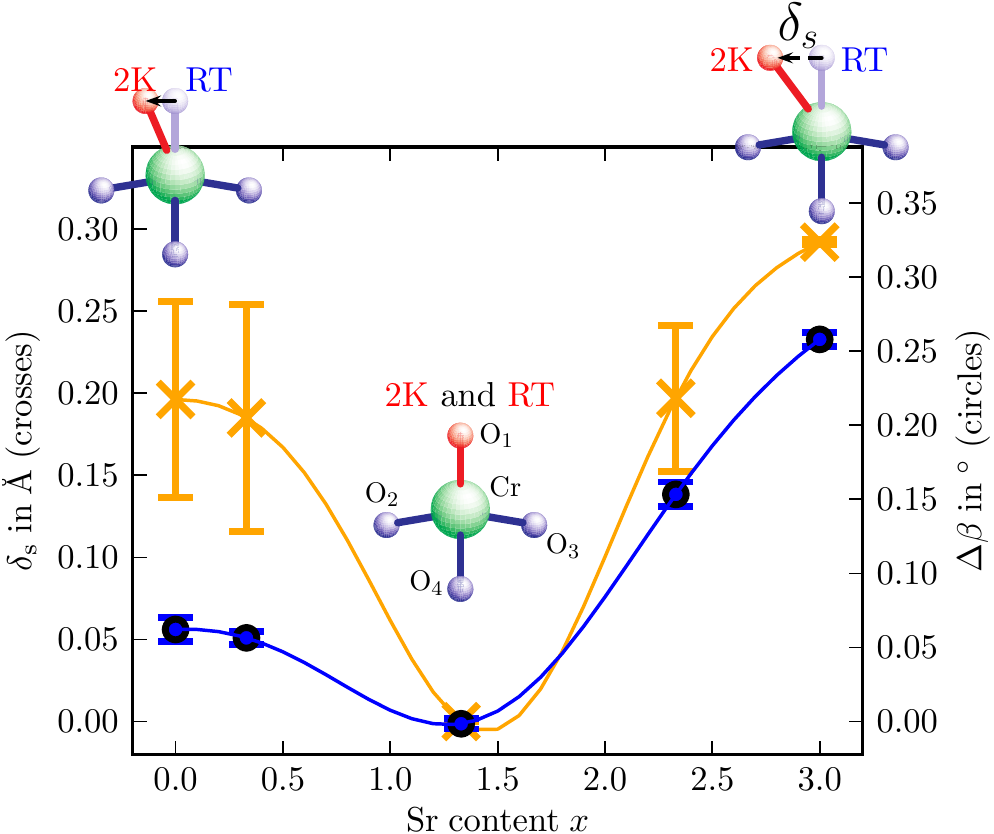}
 \caption{Distance \(\delta_s\) of the apical oxygen atom from the dimer axis (crosses) and the angle distortion \(\Delta\beta\) (circles) as a function of the Sr content \(x\). The solid lines are guides to the eye. The distance in the corresponding sketches is exaggerated for visual clarity.}
 \label{fig:shiftfig}
\end{figure}

Where detected, the symmetry breaking is due to a horizontal displacement \(\delta_s\) of the apical oxygen in the tetrahedron surrounding the Cr\(^{5+}\)-ions (\(O_1\) in our notation, see Figs. \ref{fig:strukturfig} and \ref{fig:shiftfig}). As shown in Fig. \ref{fig:shiftfig}, this \(\delta_s\) is smaller for intermediate \(x\) than for the parent compounds. By shifting O\(_1\) along the O\(_2\)-O\(_3\)-edge, the site symmetry of the Cr-ions (\(3m\) at room temperature) is lost. This oxygen shift affects the electronic orbitals of the Cr-ions. In the case of a perfect tetrahedron, the \textit{d}-orbitals are grouped into a lower lying, twofold degenerate \(e_g\) state and a higher lying, threefold degenerate \(t_g\) state. When the symmetry breaking occurs in \(\mathrm{Ba}_{3-x}\mathrm{Sr}_x\mathrm{Cr_2O_8}\), the \(e_g\) state degeneracy is lifted, with a separation into a lower lying \(3z^2-r^2\) orbital and a higher lying \(x^2-y^2\) orbital. Therefore, both of the \(e_g\)-states have to be considered when calculating \(\langle\Delta e\rangle\) in case of the space group R\(\overline{3}\)m, whereas only the \(3z^2-r^2\) orbital has to be taken into account for C\(_{2/c}\). 

Apart from a shift of the apical oxygen atom, the structural distortion also affects the monoclinic angle \(\beta\). Where the distortion could be detected, \(\beta\) deviates slightly by \(\Delta\beta\) from the value that is expected from a conversion between the rhombohedral and the monoclinic space group (see Fig. \ref{fig:strukturfig}). The value of \(\Delta\beta\) was found to depend on \(x\) (see Fig. \ref{fig:shiftfig}) and vanishes for stoichiometries where the Jahn-Teller transition is suppressed. 

\subsection*{Calculating \(J_0\) from the structural data}
To quantitatively examine the effect of the oxygen shift on the interaction constant \(J_0\), we calculated the energy splitting \(\langle\Delta e\rangle\) for the obtained crystals structures and estimated the repulsion potential \(U\). Unfortunately, no experimental values for \(J_0\) at room temperature have been reported. To obtain reasonable values for \(U\), we therefore first carried out ETHB calculations based on the low temperature structures for \(x=0\) and \(x=3\). For the occupied \(3z^2-r^2\), our calculations calculations yielded orbital splitting energies \(\Delta e(x=0)=5.6\, \text{meV}\) and \(\Delta e(x=0)=3.9\, \text{meV}\). From these values and the reported interaction constants \(J_0^\mathrm{Ba}=2.38\,\mathrm{meV}\)\cite{kofu} and \(J_0^\mathrm{Sr}=5.55\,\mathrm{meV}\)\cite{quinteracastro_sr3cr2o8}, we obtained \(A=-216\,\text{meV}\) and \(B=648\,\text{meV}\). Based on these parameters, we could estimate the value of repulsion potential \(U\) for all examined values of \(x\). As we assumed \(U\) to be temperature independent, the same values of \(U(x)\) were used to calculate for \(J_0\) at low temperature and at room temperature from the calculated splittings \(\langle\Delta e\rangle\).

\begin{figure}
 \includegraphics{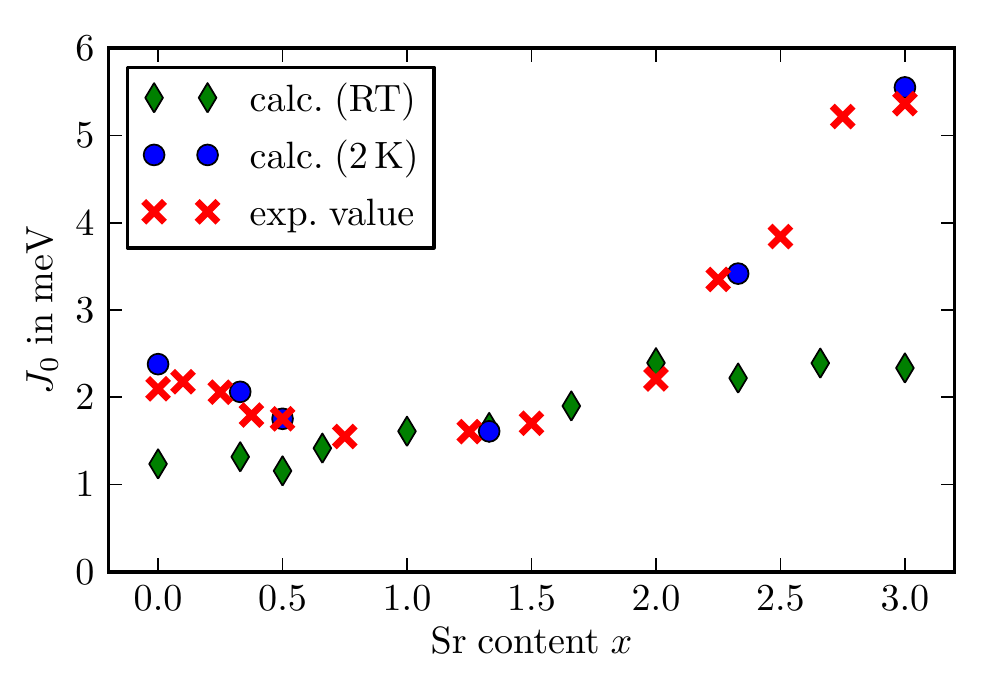}
 \caption{Intradimer interaction constant \(J_0\) for \(\mathrm{Ba}_{3-x}\mathrm{Sr}_x\mathrm{Cr_2O_8}\) as a function of the Sr content \(x\). The shown data points are experimentally obtained from SQUID-measurements and estimated based on EHTB-calculations for the crystal structure at room temperature and \(T=2\,\mathrm{K}\), respectively.}
 \label{fig:ww-konstanten}
\end{figure}

In  Fig. \ref{fig:ww-konstanten}, we have plotted the previously reported\cite{mrbpaper} values of \(J_0\) as obtained from fitting magnetization data \(M(T)\) together with the results of our calculations of \(J_0\) at room temperature and at \(T=2\,\mathrm{K}\). The experimental values of \(J_0\) are most sensitive to the low-temperature magnetization part of the \(M(T)\) data used for the fitting procedure. Therefore, they are valid well below room temperature. Fig. \ref{fig:ww-konstanten} demonstrates, as expected,  that the experimental values can be well reproduced within the ETHB framework if the corresponding calculations are based on the low temperature structure. By contrast, the calculated room temperature values of \(J_0\) underestimate the experimental values for a wide range of \(x\). On the other hand, the experimental values and the calculations for both the room temperature and low temperature structure almost coincide for a Sr content of \(x\approx 1.33\). For this composition, our neutron diffraction experiments did not indicate any evidence for the structural symmetry breaking that we observed for the other values of \(x\) at \(T=2\,\mathrm{K}\). Furthermore, combining the results of Figs. \ref{fig:shiftfig} and  \ref{fig:ww-konstanten}, suggests that a larger value of \(J_0\) seems to be accompanied by a stronger symmetry breaking.

From these results, we conclude that the Jahn-Teller distortion, which induces an orbital ordering \cite{whangbo_ba3cr2o8,Chapon}, increases the intradimer interaction constant \(J_0\) for \(0<x<3\), as it was reported for \(\mathrm{Sr_3Cr_2O_8}\). Without this orbital ordering, the splitting \(\langle \Delta e\rangle\) for a given \(x\) remains almost constant for all temperatures. We predict that no (or just a minimal) distortion will be present for stoichiometries for which the experimental value of \(J_0\) is close to the calculated room temperature value. 

\subsection*{Possible reasons for the suppressed Jahn-Teller distortion}

\begin{figure}
 \includegraphics{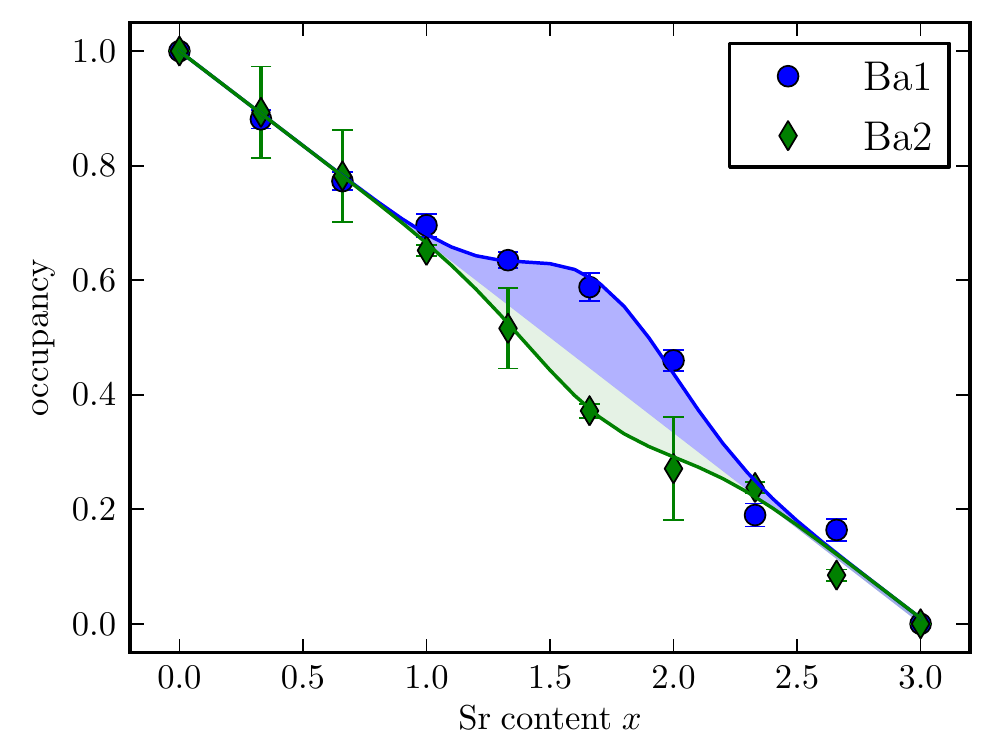}
 \caption{Obtained occupancy for the Ba atoms at sites 1 and 2 as a function of the Sr content \(x\) obtained at room temperature. The solid lines are guides to the eye, the filled areas mark the difference to the expected strictly linear decrease.}
 \label{fig:okkufig}
\end{figure}

To provide an explanation for the suppressed Jahn-Teller distortion, we analyzed the structural data obtained at room temperature in search for non-linearities as a function of the Sr content \(x\). A detailed analysis of the site occupancies revealed that the Ba/Sr-distribution over the possible sites does not directly follow the stoichiometry (see Fig. \ref{fig:okkufig}). We found that the occupation probability for Ba at the Ba\(_1\) site at intermediate stoichiometries is enhanced with respect to a completely random case and vice versa for the Ba\(_2\) site. Albeit, the occupation preferences are not very pronounced and do not fully coincide with the suppression of the Jahn-Teller distortion. In addition, such site preferences do not directly break the site symmetry of the Cr\(^{5+}\)-ions, so that a Jahn-Teller distortion could still be energetically favorable. As all other quantities describing the average crystal structure changes linearly as a function of the Sr content \(x\)\cite{supplementary_material}, there seems to be no obvious structural reason for the suppression  of this distortion.

However, as the prerequisite for such a distortion is a global degeneracy of electronic orbitals it could also be suppressed by local changes in the crystal structure. While for pure Ba\(_3\)Cr\(_2\)O\(_8\) and Sr\(_3\)Cr\(_2\)O\(_8\), all Cr-O distances are well defined, we can expect strong local deviations from the average for intermediate values of \(x\) due to the very different ionic radii of Ba\(^{2+}\) and Sr\(^{2+}\). The different distributions of Sr/Ba-ions inside the unit cell should thus lead to an increasing width of the distribution of Cr-O distances\cite{Marisa_disorder}. The symmetry at the Cr-site would then be broken locally and preserved only on average. Such a symmetry breaking then lifts the degeneracy of the \(e_g\) state by an energy difference \(\Delta E(x)\). As \(\Delta E(x)\) increases, the possible energy gain of the phase transition is lowered and therefore this transition itself is gradually suppressed.

\begin{figure}
 \includegraphics{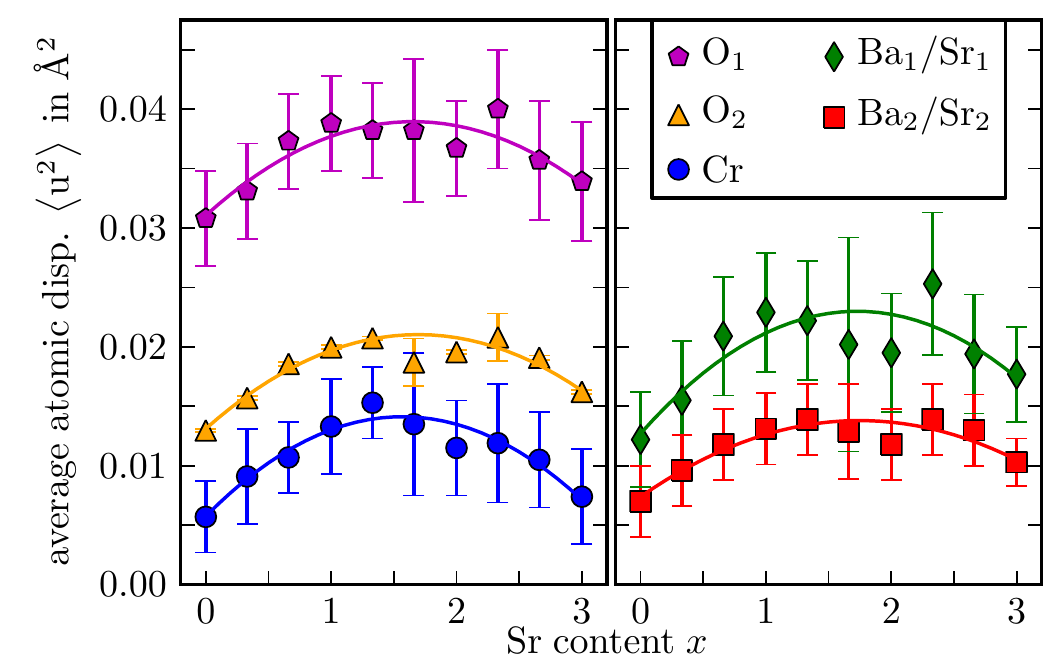}
 \caption{Atomic displacements \(\left\langle u^2\right\rangle\) in \(\mathrm{Ba}_{3-x}\mathrm{Sr}_x\mathrm{Cr_2O_8}\) at room temperature as a function of the Sr content \(x\).}
 \label{fig:debw_parted}
\end{figure}

\begin{figure}
 \includegraphics{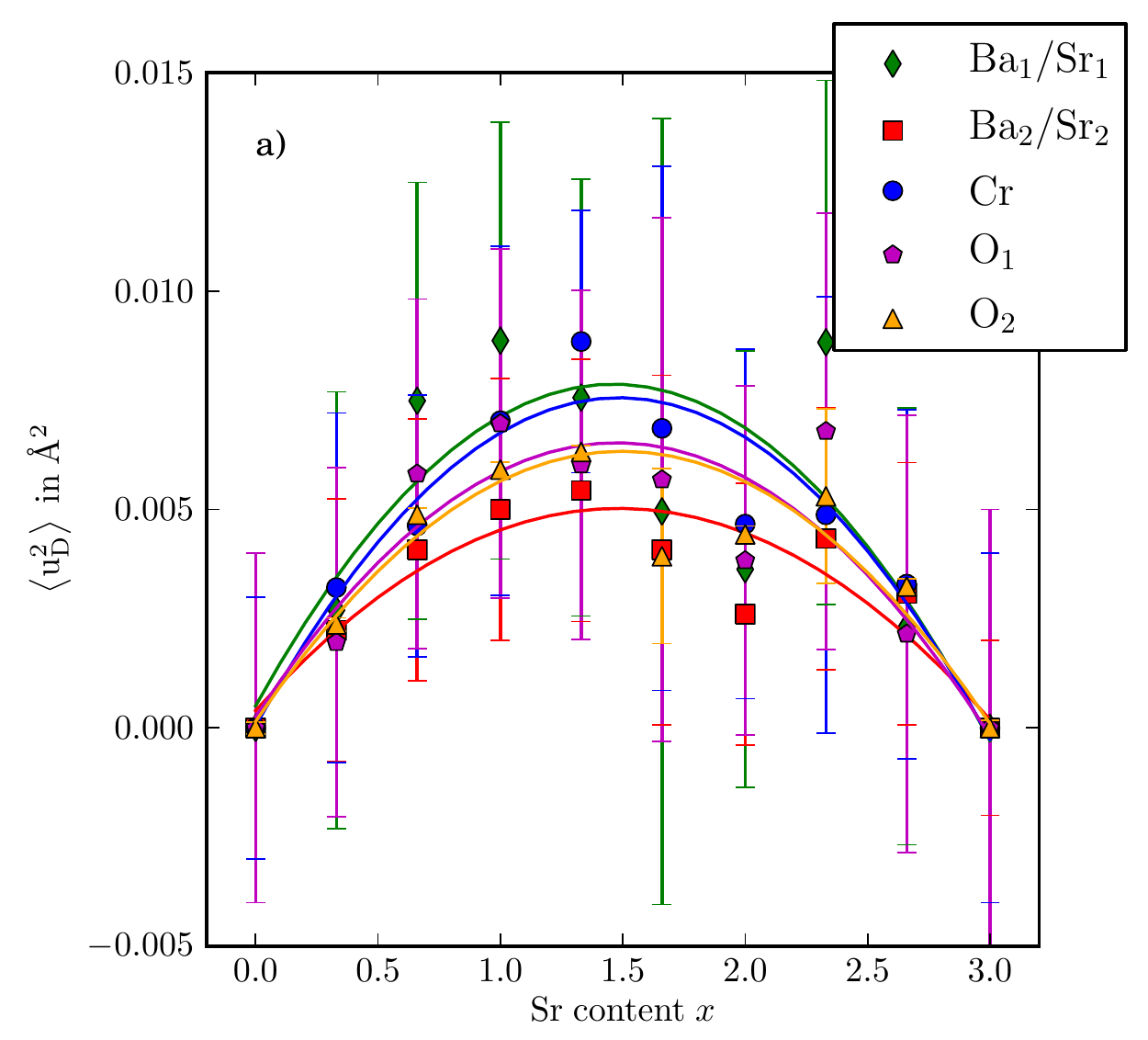}\\
 \includegraphics{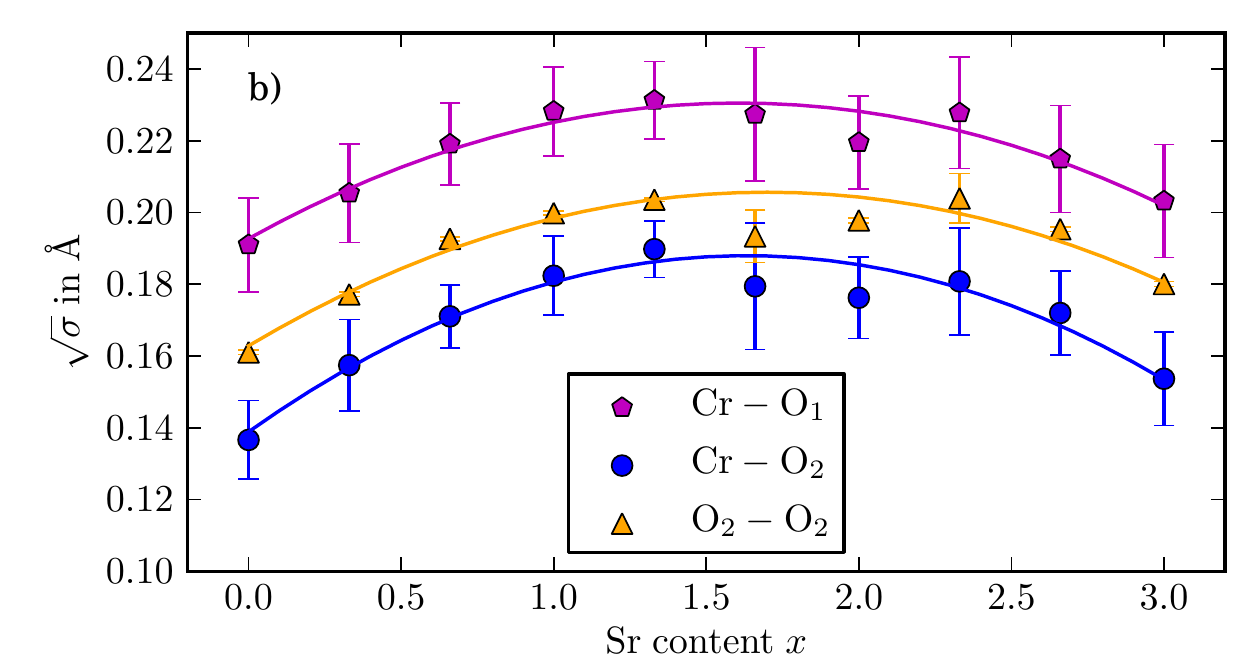}
 \caption{a): Estimated disorder contribution \(\left\langle u_\text{D}^2\right\rangle\) to the atomic displacements \(\left\langle u^2\right\rangle\), in \(\mathrm{Ba}_{3-x}\mathrm{Sr}_x\mathrm{Cr_2O_8}\) at room temperature as a function of the Sr content \(x\). b): Square root of the variance of atomic distances relevant to the electronic energies in the oxygen tetrahedron surrounding the Cr-ions and for the interaction constant \(J_0\). The solid lines are guides to the eye.}
 \label{fig:debyewaller_subtract}
\end{figure}

This scenario seems to be confirmed by the variation of the mean square displacements \(\left\langle u^2\right\rangle\) as deduced from the structural refinement and shown in Fig. \ref{fig:debw_parted}, which are found to be maximum for intermediate values of \(x\). We attribute this behavior to an increase of the local disorder as it is common in for solid solutions. To obtain an estimate for this disorder contribution, we have assumed that \(u\) is given by Gaussian distributions and comprises a thermal and a disorder term, \(u=u_T+u_D\), with no correlation between \(u_T\) and \(u_D\). Therefore, the total mean displacement would be \(\left\langle u^2\right\rangle=\left\langle u_T^2\right\rangle+\left\langle u^2_D\right\rangle+2\left\langle u_Du_T\right\rangle=\left\langle u_T^2\right\rangle+\left\langle u^2_D\right\rangle\). We have assumed the thermal part to be linear in \(x\), \(\left\langle u_T^2\right\rangle(x)=\rho x+\tau\), and calculated the corresponding values for \(\rho\) and \(\tau\) based on \(\left\langle u^2\right\rangle\) for \(x=0\) and \(x=3\). We subtracted this thermal part from the experimentally obtained displacements and plotted the resulting \(\left\langle u_D^2\right\rangle(x)\) in  Fig. \ref{fig:debyewaller_subtract}b. 
Furthermore, as our argument mainly concerns variations of the distances between the oxygen and the chromium atoms, we have estimated the square root of the variance \(\sqrt{\sigma}_d\) of the atomic distances \(d\) inside the tetrahedron. The value of \(\sqrt{\sigma}_d=\sqrt{\left\langle(d-\left\langle d\right\rangle)^2\right\rangle}\) is derived from the mean displacements and gives a more concrete picture of the influence of the chemical disorder on the distances which are important for the interaction constant and the orbital energies of the Cr-atom that drive the Jahn-Teller distortion (see Fig. \ref{fig:debyewaller_subtract}a). All these disorder indicators peak where the structural transition is suppressed. We thus believe that this type of disorder is most likely at the origin of the suppression of the Jahn-Teller induced structural phase transition. 

\section*{Summary and Outlook}
We have performed  neutron powder diffraction experiments to obtain detailed structural information  about \(\mathrm{Ba}_{3-x}\mathrm{Sr}_x\mathrm{Cr_2O_8}\) for various values of \(x\) at room temperature and \(T=2\,\mathrm{K}\). Based on these diffraction experiments, we have shown that the Jahn-Teller distortion reported for the parent compounds is present in \(\mathrm{Ba}_{3-x}\mathrm{Sr}_x\mathrm{Cr_2O_8}\), but seems to be gradually suppressed for intermediate values of \(x\).  We have given an explanation for this suppressed symmetry breaking at intermediate values of \(x\) based on an increasing disorder in the system. To test our hypothesis, experiments to directly probe the Cr-states as a function of the Sr content \(x\) would be extremely helpful. Furthermore, DFT-calculations of the total energy per unit cell for the two possible space groups, similar to \cite{sr3cr2o8_dft}, could clarify whether or not any symmetry breaking should be present in a disorder free system with the respective equilibrium crystal structures for intermediate values of \(x\).  We have demonstrated that the interdimer interaction constant \(J_0\) can be calculated within the EHTB framework. We furthermore showed that the reported, peculiar change of \(J_0\) can be explained based on the varying strength of this Jahn-Teller distortion. The tuning of \(J_0\) with varying \(x\) as reported here should, along with a corresponding variation of the other relevant interaction constants, allow for a direct control of the critical fields \(H_c\) in the \(\mathrm{Ba}_{3-x}\mathrm{Sr}_x\mathrm{Cr_2O_8}\) system.

%
\end{document}